\documentclass[aps,pra, twocolumn, superscriptaddress,amssymb,showpacs]{revtex4}
\usepackage{graphicx}
\usepackage{epstopdf}
\begin{document}
\title{Multiphoton blockades in pulsed regimes beyond the stationary limits}

\author{G.~H.~ Hovsepyan}
\email[]{gor.hovsepyan@ysu.am}
\affiliation{Institute for Physical Researches,
National Academy of Sciences,\\Ashtarak-2, 0203, Ashtarak,
Armenia}\affiliation{Yerevan State University, Alex Manoogian 1, 0025,
Yerevan, Armenia} 

\author{A.~R.~Shahinyan}
\email[]{anna\_shahinyan@ysu.am}
\affiliation{Institute for Physical Researches,
National Academy of Sciences,\\Ashtarak-2, 0203, Ashtarak, Armenia}

\author{G.~Yu.~Kryuchkyan}
\email[]{kryuchkyan@ysu.am}
\affiliation{Institute for Physical Researches,
National Academy of Sciences,\\Ashtarak-2, 0203, Ashtarak,
Armenia}\affiliation{Yerevan State University, Alex Manoogian 1, 0025,
Yerevan, Armenia}
\pacs{42.50.Dv, 42.50.Gy, 42.50.Lc}

\begin{abstract}
{We demonstrate multiphoton blockades (PB) in the pulsed regime by using Kerr nonlinear dissipative resonator driven by a sequence of Gaussian pulses. It is shown that the results obtained for single-photon, two-photon and three-photon blockades in the pulsed excitation regime differ
considerably from analogous results obtained for the case of continuous-wave
(cw) driving. We strongly demonstrate that for the case of cw pumping of the
Kerr-nonlinear resonator there are fundamental limits on populations of lower
photonic number-states (with $n = 0, 1, 2, 3$). 
Thus, such detailed comparison demonstrates that PB due to excitation with a suitable photon pulses is realized beyond the fundamental limits established for cw excitations. We analyze photon-number effects and investigate phase-space properties of PB
on the base of photon number populations, the second-order correlation functions and the Wigner functions in phase space. Generation of Fock states due to PB in the pulsed regime is analysed in details.}
\end{abstract}

\pacs{}

\maketitle

\section{Introduction}

The realization of quantum devices with photon-photon interaction has become an interesting and important research topic in quantum optics and quantum technologies. The key element to obtain the mentioned interactions is strong optical nonlinearity that gives rise to many important quantum effects at level of few-photons, including photon blockade\cite{Imam, Wer, Birn}.

 In the photon blockade (PB) the optical response to a single photon is modulated by
the presence or absence of the other photons. Particularly, the capture of a single
photon into the system affects the probability that a second photon is admitted. Thus, the photon blockade is the analog of the Coulomb blockade for electrons in condensed matter devices, where single electron transport is blocked
by a strong Coulomb interaction in a confined structure. A simple consequence of photon blockade is the antibunching 
of photons in emission in analogy to the photon antibunching of resonance fluorescence on a two-level atom \cite{Carm, Kim}.

 One of
the basic conditions for PB is that the photon-photon interaction strength should be larger regarding decay rate of the system. In this respect, strong nonlinearities on the few-photon level can be produced by interaction between photons and an atom in a cavity \cite{Birn2, Bish, Fink, Far}, in systems with interacting photons or polaritons in
arrays of cavities coupled to atoms or qubits \cite{Hart, Utus, Schm, Tom}, in optomechanical systems and Kerr type nonlinear cavities \cite{Lia, Ferr, Mir}.

Photon blockade was first
observed in an optical cavity coupled to a single trapped
atom \cite{Birn}. The PB has been predicted in cavity quantum
electrodynamics (QED) \cite{Tia}, and recently in circuit
QED with a single superconducting artificial atom coupled to a microwave transmission line
resonator \cite{Hof,Lan}. PB was also experimentally demonstrated with a photonic crystal cavity with a strongly coupled quantum dot \cite{Far2},
and was also predicted in quantum optomechanical systems \cite{Rab, Nun}. An analogous phenomenon of phonon blockade was predicted for an artificial superconducting atom coupled to a nanomechanical resonator \cite{Liu}, as well as 
the polariton blockade effect due to polariton-polariton interactions has been considered in \cite{Ver}. Recently, PB was considered in dispersive qubit-field interactions in a superconductive coplanar waveguide cavity \cite{hof2} and with time-modulated input \cite{A}.

In most of the cited works PB was considered in nonlinear optical systems
driven by a continuous-wave laser. However, photon sources on demand require
production of photon pulses at fixed time-intervals. In this paper, we
investigate PB in the pulsed regime, considering Kerr-type oscillatory
dissipative system driven by a sequence of classical Gaussian pulses separated
by time intervals. This approach allows to expand the mechanism of PB for the
case of producing few-photon pulses at deterministic time intervals. Thus,
we present comparative analysis of one-photon, two-photon and three-photon
blockades in the pulsed regimes in addition to the results obtained for the
case of PB in cw driving field \cite{Mir}. 

In this way, we demonstrate, that contrarily to common expectations, the results obtained in the pulsed regime considerably differ from those derived for the monochromatic driving \cite{Mir}.
More importantly, we see that PB in Kerr-type systems under pulsed excitation can be controlled by shape of pulses. In this spirit, we emphasize the idea of improving the degree of quantum effects in open systems, 
as well as obtaining qualitatively new quantum effects by applying of the sequence of tailored pulses. This approach was recently exploited for formation of high degree continuous-variable entanglement in the nondegenerate optical parametric
oscillator \cite{adam, adam2}, and for investigation of quantum interference
in mesoscopic domain \cite{gev1, gev2, gev3}. Recently, generation of
nonclassical states in cavity QED with pulsed input has been investigated in
\cite{A} and production of a superposition states in the periodically pulsed
nonlinear resonator has been demonstrated in \cite{gev2}, \cite{gev4}.  Dynamics of
periodically driven nonlinear oscillator has been also studied in the papers
\cite{AID}-\ \cite{LM}.

We clarify the multiphoton photon blockade in Kerr nonlinear resonator under pulsed excitation mainly by considering photon-number effects and by analysing phase-space properties of resonator mode. However, we have not discussed properties of the field transmission when the photons are blockaded. 
Thus, we focus on analysis of the mean photon number, the probability distributions of photons, the second-order correlation functions of photons and the Wigner functions in phase space. In this pulsed regime the ensemble-averaged mean
photon numbers, the populations of photon-number states and the Wigner functions are nonstationary and exhibit a periodic time dependent behavior, i.e. repeat the periodicity of the pump laser in an over transient regime.
Besides this the results on production of photon Fock states at PB region depend on the parameters of Gaussian pulses such as the amplitude, the duration of pulses and the time-interval between them, dissipation rates and Kerr-interaction coupling. Thus, we investigate production of photonic states for an arbitrary interaction time-intervals including also time-intervals
exceeding the characteristic time of dissipative processes, $t\gg\gamma^{-1}$. However, the results obtained are nonstationary that is conditioned by interaction of the system with the sequence of Gaussian e.m. pulses. 

For comparison, the paper includes the investigation of the Kerr nonlinear dissipative resonator under cw driving on the framework of the potential solution of the Fokker-Planck equation for the quasiprobability distribution function in complex
P-representation. In this way, we obtain that contrarily to the pulsed regime for the case of cw pumping there are fundamental limits on the populations of lower photonic-number-states.

The paper is arranged as follows. In Sec. II we describe periodically pulsed Kerr nonlinear resonator and discuss the case of cw driving in the exact quantum treatment of PB. In Sec. III we consider one-photon, two-photon and three-photon blockades in the pulsed regime on base of the populations of photon-number states and the second-order correlation functions. In Sec. IV we analyse the production of multi-photon states on base of the Wigner functions. We summarize our results in Sec. V.

\section{Kerr nonlinear resonator: pulsed regime and cw limit}

\begin{figure}
\includegraphics[width=8.6cm]{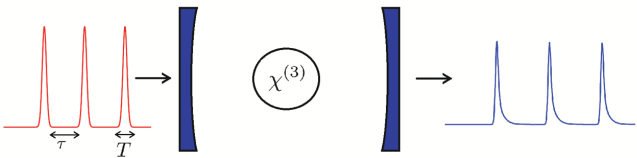}
\caption{Schematic representation of Kerr nonlinear resonator under a sequence
of Gaussian pulses. In this scheme a emitter involving third-order
susceptibility coupled to a resonator.}
\label{NRO}
\end{figure}

The Hamiltonian of Kerr nonlinear resonator under pulsed excitation in the rotating wave approximation reads as:
\begin{equation}
H=\hbar \Delta a^{+}a + \hbar \chi (a^{+})^{2}a^{2} +
\hbar f(t)(\Omega a^{+} + \Omega^{*}a).\label{Hamiltonian}
\end{equation}
Here, time dependent coupling constant $\Omega f(t)$ that is proportional to the amplitude of the driving field consists of the Gaussian pulses with the duration $T$ which are separated by time intervals $\tau$
\begin{equation}
f(t)=\sum{e^{-(t - t_{0} - n\tau)^{2}/T^{2}}}, \label{driving}
\end{equation}
while $a^{+}$, $a$ are the creation and annihilation operators, $\chi$ is the
nonlinearity strength, and $\Delta=\omega_{0} -\omega$ is the detuning
between the mean frequency of the driving field and the frequency of the
oscillator, (see Fig. \ref{NRO}).

This model seems experimentally feasible and can be realized in
several physical systems. Particularly, the effective Hamiltonian (\ref{Hamiltonian}) describes a
qubit off-resonantly coupled to a driven cavity. In fact, it is well known that the Hamiltonian of two-level atom interacting with cavity mode in the dispersive approximation, if the two-level system remains in its ground state, can be reduced to the effective Hamiltonian (\ref{Hamiltonian}). This model also describes a nanomechanical oscillator with $a^{\dagger}$ and $a$ raising and lowering operators related to the position and momentum operators of a mode quantum motion. An important implementation of Kerr-type resonator has been recently achieved in the
context of superconducting devices based on the nonlinearity
of the Josephson junction.

We have included dissipation and decoherence in Kerr nonlinear resonator on the basis of the master
equation:
\begin{equation}
\frac{d \rho}{dt} =-\frac{i}{\hbar}[H, \rho] +
\sum_{i=1,2}\left( L_{i}\rho
L_{i}^{\dagger}-\frac{1}{2}L_{i}^{\dagger}L_{i}\rho-\frac{1}{2}\rho L_{i}^{\dagger}
L_{i}\right)\label{master},
\end{equation}
where $L_{1}=\sqrt{(N+1)\gamma}a$ and $L_{2}=\sqrt{N\gamma}a^+$ are the
Lindblad operators, $\gamma$ is a dissipation rate, and $N$ denotes the mean
number of quanta of a heat bath.
To study the pure quantum effects, we focus on the cases of very low reservoir temperatures, which, however, ought to be still
larger than the characteristic temperature $T \gg T_{cr}=\hbar\gamma/k_B$. This
restriction implies that dissipative effects can be described self-consistently
in the frame of the Linblad Eq. (\ref{master}). In our numerical
calculation we choose the mean number of reservoir photons $N=0.01$. 

Now we describe operational regimes of the system that are relevant to multiphoton blockades.
In the absence of any driving, states of the Hamiltonian (\ref{Hamiltonian})
are the Fock photon states $|n\rangle$ which are spaced in energy $E_{n} = E_{0} +
\hbar\omega_{0} n + \hbar\chi n(n-1)$ with $n = 0, 1, ...$. The
levels form an anharmonic ladder with
anharmonicity that is given by $E_{21}-E_{10}=2\hbar\chi$.

\begin{figure}
\includegraphics[width=8.6cm]{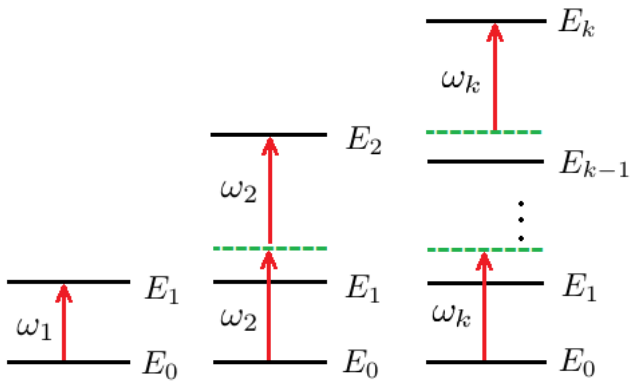}
\caption{Schematic energy-level diagram with one-photon, two-photon and $ k$
-photon resonant transitions between states of Kerr nonlinear resonator.
Selective excitations leading to multiphoton blockades are realized for the main
frequencies of driving field: $ k\hbar \omega_k=E_{k0}$ that are calculated as
$\omega_k=\omega_0+\chi (k-1)$. }
\label{energyL}
\end{figure}

Below, we concentrate on quantum regimes
for the parameters leading to resolved oscillatory energy levels. In this case we can consider near the resonant selective transitions between lower photon-number
states $|m\rangle\rightarrow|n\rangle$. In this way, $ k$-photon blockades for the positive $k=0, 1, 2, 3,...$ are realized in both resonant and near the resonant transitions between oscillatory initial and final states $|0\rangle\rightarrow|k\rangle$. According to the formula $E_{n} = E_{0} +
\hbar\omega_{0} n + \hbar\chi n(n-1)$ the resonant frequencies of these transitions equal to $ k\hbar \omega_k=E_{k0}$ and can be derived in the following form $\omega_k=\omega_0+\chi (k-1)$ (see, Fig. \ref{energyL}).
Thus, for one-photon transition $k=1$ and $E_{10}=\hbar \omega_0$, the resonance frequency is $\omega_1=\omega_0$,
for two-photon transition $k=2$ and $E_{20}=2 \hbar \omega_0 + 2 \chi $, the other resonance frequency is $\omega_2=\omega_0+ \chi $, while for $k=3$ , $E_{30}=3 \hbar \omega_0 + 6 \chi $, the resonance frequency is $\omega_3=\omega_0+ 2\chi $.

Considering the pulsed regimes of Kerr nonlinear reservoir we assume that the
spectral pulse-width, i.e. the spectral widths of pulses, should be smaller
than the nonlinear shifts of the oscillatory energy levels. It means that the
duration of pulses should be larger than $1/\chi $. Thus, for strong
nonlinearities 
$\chi/\gamma >1$, we arrive to the following inequalities for the duration of Gaussian pulses $1/\gamma > T > 1/\chi$.

We solve the master equation Eq. (\ref{master}) numerically based on quantum
state diffusion method.  The applications of this method for studies of NDO
can be found in \cite{gev1}-\cite{gev4}.
In the calculations, a finite basis of number states $|n\rangle$ is kept large
enough (where $n_{max}$ is typically 50) so that the highest energy states are
never populated appreciably. In the following the distribution of oscillatory
excitation states
$P(n)=\langle n|\rho|n\rangle$ as well as the Wigner functions
\begin{equation}
W(r, \theta)=\sum_{n,m}\rho_{nm}(t)W_{mn}(r,\theta)
\label{expr:wigner}
\end{equation}
in terms of the
matrix elements $\rho_{nm}=\langle n|\rho|m\rangle$ of the density
operator in the Fock state representation
will be calculated.
Here $(r,\theta)$ are the polar coordinates in the complex
phase-space plane, $x=r\cos\theta$, $y=r\sin\theta$, whereas the
coefficients $W_{mn}(r,\theta)$ are the Fourier transform of
matrix elements of the Wigner characteristic function.

In this way, we calculate below nonstationary representation of a single
emitter coupled to an optical resonator populations $P_0, P_1, P_2, P_3$ of photon-number states $|n\rangle$, $n=0, 1, 2, 3$, as well as the mean photon number, the Wigner functions and the second-order correlation functions.

\subsection{ Limits to the state populations for the case of cw driving}

In this subsection we shortly discuss selective excitation of a Kerr nonlinear
resonator under cw driving. In this case the Hamiltonian (\ref{Hamiltonian})
reads as:
\begin{equation}
H=\hbar \Delta a^{\dagger}a + \hbar \chi (a^{\dagger})^{2}a^{2} +
\hbar (\Omega a^{\dagger} + \Omega^{*}a),\label{hamiltonian}
\end{equation}
i.e. describes a standard anharmonic oscillator driven by a monochromatic field in the rotating wave approximation. In this simplest case,
an analytical results for a dissipative driven nonlinear oscillator in steady state have
been obtained in terms of the solution of the Fokker-Planck equation for the quasiprobability distribution function $P(\alpha, \alpha^{*})$ in complex
P-representation \cite{drum}. This approach based on the
method of potential equations leads to the analytic solution for the quasiprobability distribution function $P(\alpha,  \alpha^{*})$ within the framework
of an exact nonlinear treatment of quantum
fluctuations. In this way, the photon number probability distribution function 
$p(n) = \langle n |\rho| n \rangle$ can be expressed in terms of complex
$P$-representation as follows:
\begin{equation}
\label{pnPc}
p(n)=\frac{1}{n}\int \int_C d\alpha d\alpha\dagger (\alpha)^n (\alpha\dagger)^nexp(-\alpha \alpha\dagger)P(\alpha, \alpha\dagger),
\end{equation}
where C is an appropriate integration contour for each of
the variables $\alpha$ and $\alpha^{*}$, in the individual complex planes.
After integration the mean photon number and the probability distribution of
photon-number states are represented in the following form \cite{a33,a34,kh}:
\begin{equation}
\langle a^{\dagger}a\rangle = \frac{\Omega^2}{(\Delta + \chi)^2 + (\gamma/2)^2}
\frac{F(\lambda + 1, \lambda^{*} + 1, z)}{F(\lambda,\lambda^{*}, 2 |\varepsilon|^{2}|)},
\label{quant}
\end{equation}
where $\varepsilon = \Omega/\chi$, $\lambda = (\gamma + i\Delta)/i\chi$, and $F = _0F_{2}$ is the hypergeometric function:
\begin{equation}
_0\left.F_2\right(a, b, z)=\sum_{k = 0}^{\infty}\frac{z^{k}\Gamma(a)\Gamma(b)}{k!\Gamma(k+a)\Gamma(k+b)},
\end{equation}

\begin{equation} 
P(n) = \frac{|\varepsilon|^{2n}\Gamma(\lambda)\Gamma(\lambda^{*})}{n!_{0}F_2(\lambda, \lambda^{*}, 2 |\varepsilon|^{2}|)}\sum_{k = 0}^{\infty}\frac{|\varepsilon|^{2k}}{k!\Gamma(k+n+\lambda)\Gamma(k+n+\lambda^{*})}\label{photon}.
\end{equation}

\begin{figure}
\includegraphics[width=8.6cm]{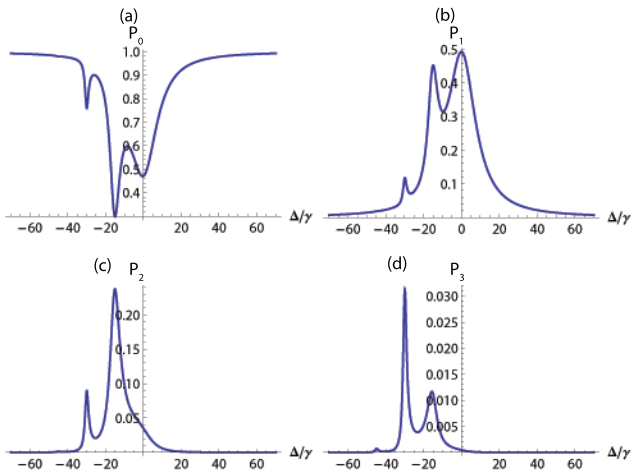}
\caption{Populations of photon number depending on the detuning. The
parameters are: $\Omega/\gamma=20$, $\chi/\gamma=15$. }
\label{monodelta}
\end{figure}

\begin{figure}
\includegraphics[width=8.6cm]{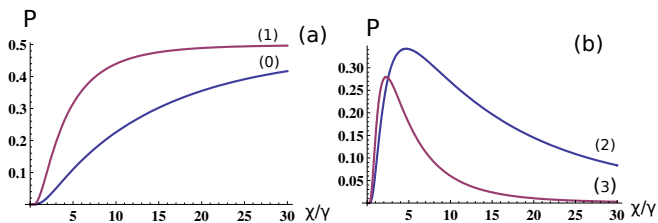}
\caption{Populations of photon numbers depending on the parameter of nonlinearity: the curves of $P_0, P_1$ populations (a), the curves of $P_2, P_3$ populations (b).
The parameters are: $\Omega/\gamma=20$, $\Delta/\gamma=0$. }
\label{monochi}
\end{figure}

In Fig. \ref{monodelta} we present the typical results for the populations
$P_0, P_1, P_2, P_3$ of the photon-number states $|n\rangle$ versus the
detuning that are calculated on the Eq. (\ref{photon}).
These results show the selective excitations of the Fock states $|1\rangle$,
$|2\rangle$, $|3\rangle$. As we see, the population $ P_1$ depicted in Fig.
\ref{monodelta}(b) displays the maximum 0.5 at $\Delta=0$, that corresponds to
the resonance
transition $|0\rangle \rightarrow |1\rangle$ at the frequency $\omega = \omega_{0}$. 
The other maximum $ P_1 = 0.46$ at $\Delta/\gamma=-15$ corresponds to the
population of $|1\rangle$ state through the Raman process with the energy
conversation $E_0+2\omega_2= \omega_k+E_1$. This process
involves the excitation of Fock state $|2\rangle$, (in the transition
$|0\rangle \rightarrow  |2\rangle$ at the frequency of pump field $\omega_2=\omega_0+ \chi
$), and the decay $|2\rangle \rightarrow |1\rangle$ at frequency $\omega_k= \omega_0+
2\chi $. Note, that the population of vacuum state is zero in this spectral
range (see, Fig. \ref{monodelta}(a)). 
The population $ P_2$ depicted in Fig. \ref{monodelta}(c) displays the
maximum 0.24 at $\Delta=-15$, that corresponds to the resonance transition
$|0\rangle \rightarrow |2\rangle$ at the frequency $\omega = \omega_{0}+ \chi
$.  
As we see, the population $ P_3$ is very small for all values of the detuning.
Nevertheless, the population display maximum at $\Delta/\gamma=-30$. This
peak corresponds to three quanta excitation of $|3\rangle$ state at the
frequency $\omega_3=\omega_0 + 2\chi$.

In general, from the analytic results and numerical analysis we can conclude
that the populations are strongly limited. We demonstrate this fact in Fig.
\ref{monochi} by showing the dependence of populations from the parameter of
nonlinearity.  We observe in Fig. \ref{monochi}(a) that both populations of
vacuum and single-photon states monotonically increase with increasing of the
nonlinearity parameter.  However, the population $P_1$ is limited by the value
0.5. This effect of selective excitation can be interpreted as the
single-photon Fock state blockading the generation of two or more photons.
However, limit to $P_1$ population is restricted possibility of observation PB
in cw regime. Note, that this  result is in accordance with the numerical
results observed on framework of the  master equation \cite{Mir}. 
The behavior of the populations of two-photon and three-photon number states
depicted in Fig. \ref{monochi}(b) differ from $P_0$ and $P_1$. We observe
that these populations display peaked structures and the maximum are realized
for the definite parameters of nonlinearity. 

One of the questions that we will try to answer below is whether or not the
populations $P_n$ of $|n\rangle$ states for Kerr nonlinear dissipative
resonator driven by a sequence of Gaussian pulses can exceed the cw limits.

\section{PB in the pulsed regimes}

Now we present the results of this paper concerning selective excitations of
photon-number states and hence the observation of an effective PB due to pulsed
excitation. In this sense, we note the main peculiarity of our paper. We
investigate photon-number aspects of PB in nonstationary regimes, for an
arbitrary interaction time intervals, that also can exceed the characteristic
time of dissipative processes, $t\gg\gamma^{-1}$. These nonstationary regimes
are conditioned by the specific form of excitation. 

If the Kerr nonlinear resonator is driven by a sequence of pulses, its behavior
is modified and essentially depends from the duration of pulses as well as
time-intervals between them. In this way, we consider three important regimes
leading to one-photon, two-photon and three-photon blockades. In these regimes
the production of single-photon, two-photon- and three-photon states blockade
the generation of the other photons.
Each of these regimes are realized for the appropriate choosing parameters,
the detuning $\Delta$, the nonlinearity parameter $\chi$, the pump amplitude
$\Omega$ and the parameters of pulses. Note, that in the case of pulsed excitation the ensemble-averaged mean
oscillatory excitation numbers and the populations of oscillatory states are nonstationary and exhibit a periodic
time dependent behavior, i.e. repeat the periodicity of
the pump laser in an over transient regime. Below we demonstrate that the concrete PB regimes can be effectively prepared if the pulsed
excitation is tuned to the corresponding resonance transitions. For
convenience, we shall refer to $k$ as a tuning continuous parameter instead of
the detuning $\Delta$. Thus, by choosing the concrete value of tuning
parameter $k$ in the expression $\omega_k=\omega_0+\chi (k-1)$ we determine
the detuning.

The other conditions to improve the effectiveness of PB concern to choosing the controlling parameters of the
pulses. We assume the regimes involving short pulses separated by long time
intervals that allow to increase the weights of photon-state populations.

\begin{figure}
\includegraphics[width=8.6cm]{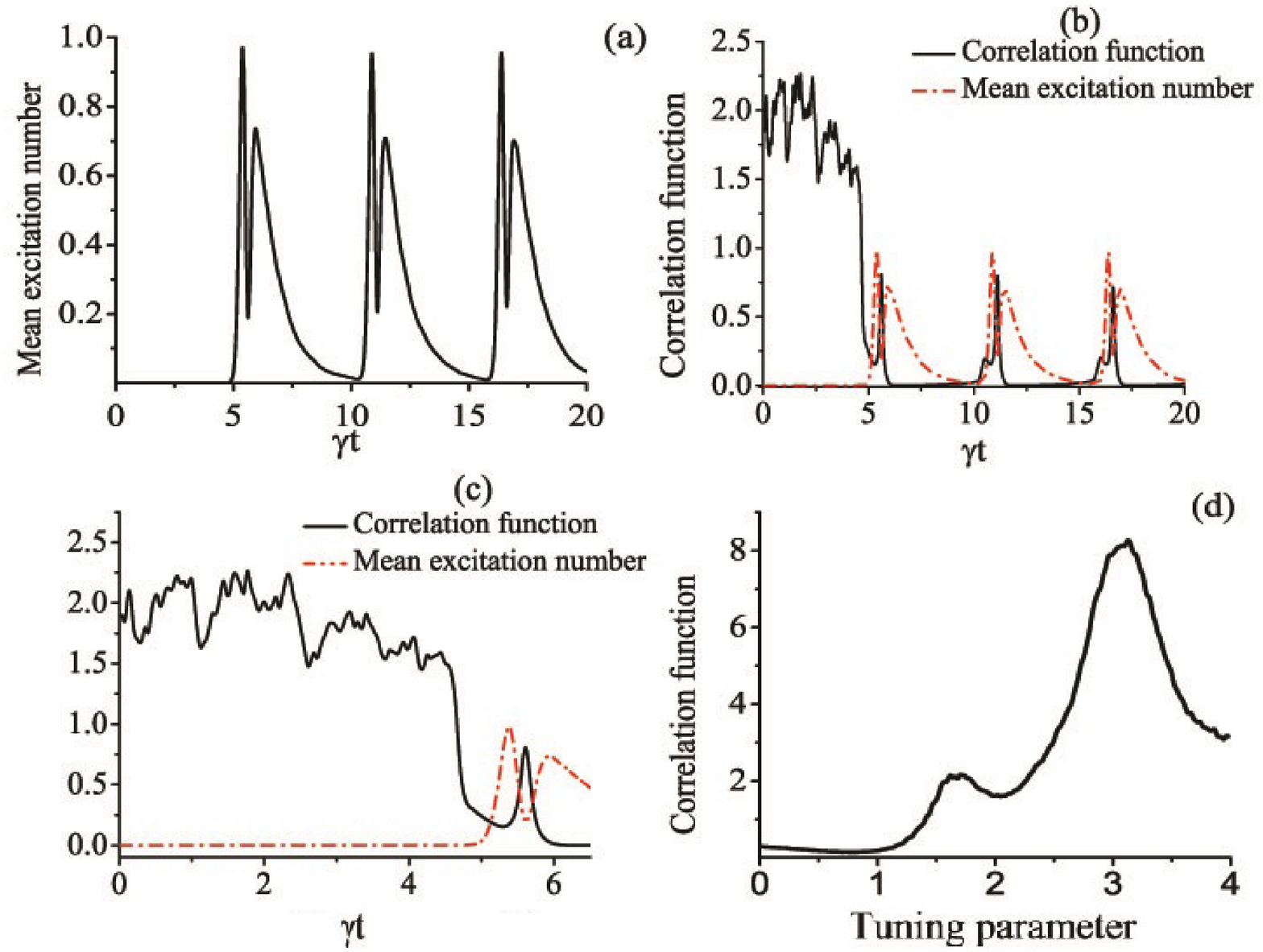}
\caption{The regime of single-photon blockade. The mean photon number versus time-interval that repeats the Gaussian pulses and also indicates decay effects, for $k=1$ (a); the time-evolution of the second-order photon correlation function versus the dimensionless time, for $k=1$ (b),(c); the second-order photon correlation function, for time-intervals at the peak values of the mean photon numbers, versus the tuning parameter (d). The parameters are:
$\chi /\gamma = 15$, the maximum amplitude of the driving field $\Omega/\gamma = 6$, the mean number of reservoir photons $N=0.01$,
$\tau = 5.5{\gamma}^{-1}$, $T=0.4{\gamma}^{-1}$.}
\label{spb}
\end{figure}

\begin{figure}
\includegraphics[width=8.6cm]{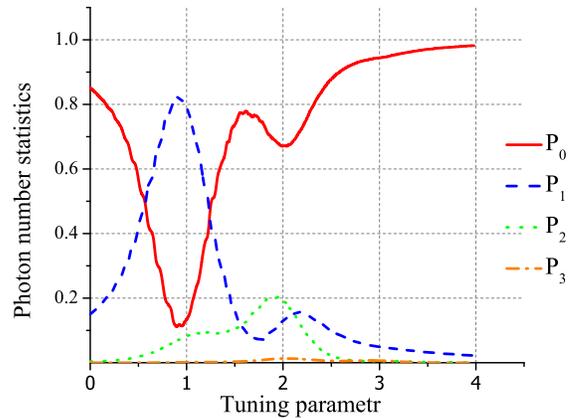}
\caption{The regime of single-photon blockade. The populations $P(n)$ of $|n\rangle$ Fock states , for $n=0, 1, 2, 3$, at the maximal peak values of the mean photon numbers, versus the tuning parameter. The parameters are:
$\chi /\gamma = 15$, the maximum amplitude of the driving field $\Omega/\gamma = 6$, the mean number of reservoir photons $N=0.01$,
$\tau = 5.5{\gamma}^{-1}$, $T=0.4{\gamma}^{-1}.$}
\label{pns}
\end{figure}

\subsection{One-photon blockade}

The analysis of the master equation Eq. (\ref{master}) shows single-photon blockade in Fig. \ref{spb} and Fig. \ref{pns}.
In this regime, the probability distribution of excitation numbers $P(n)$
displays maximum, approximately equals to unity for single-photon state, $n=1$,
for the tuning parameter $k=1$ or the detuning $\Delta=0$. We also analyze the
populations of photonic states for the definite time-intervals
corresponding to the maximal peak values of the mean photon numbers
according to Fig. \ref{spb}(a). The values of populations at these peaks versus the
tuning parameter are depicted in Fig. \ref{pns}. As we see, for the parameters: 
$\chi /\gamma = 15$, $\Omega/\gamma = 6$,
$\tau = 5.5{\gamma}^{-1}$, and $T=0.4{\gamma}^{-1}$ the maximum of $|1\rangle$
Fock state population reaches to $P_1=0.82$ approximately for $k=1$, i.e. for the near to resonance
frequency $\omega_1=\omega_0$. In this case the populations of the other
states are very small, particularly, $P_0$ of vacuum state is approximately
equals to 0.14 and  $P_2$ of $|2\rangle$ Fock state is 0.04.

It is interesting to compare these results with the solutions obtained for the
Kerr nonlinear resonator with cw driving field. It has been shown numerically
\cite{Mir} that in the case of cw-excitation the maximum rate of the population
$P_1$ reach only to 0.5 for $k=1$ and approximately equals to the population of
vacuum state $P_0$. The limit 0.5 for $P_1$ has been also obtained by
analytical calculations in the previous section. Thus, the population
$P_1=0.82$ calculated in the pulsed regime essentially exceeds the analogous
result in cw regime. This result indicates the efficiency of the single-photon
blockade in the pulsed regime. Indeed, we can conclude that such high
population of single-photon state effectively blockades the entering
of the other photons in the resonator. 

\begin{figure}
\includegraphics[width=8.6cm]{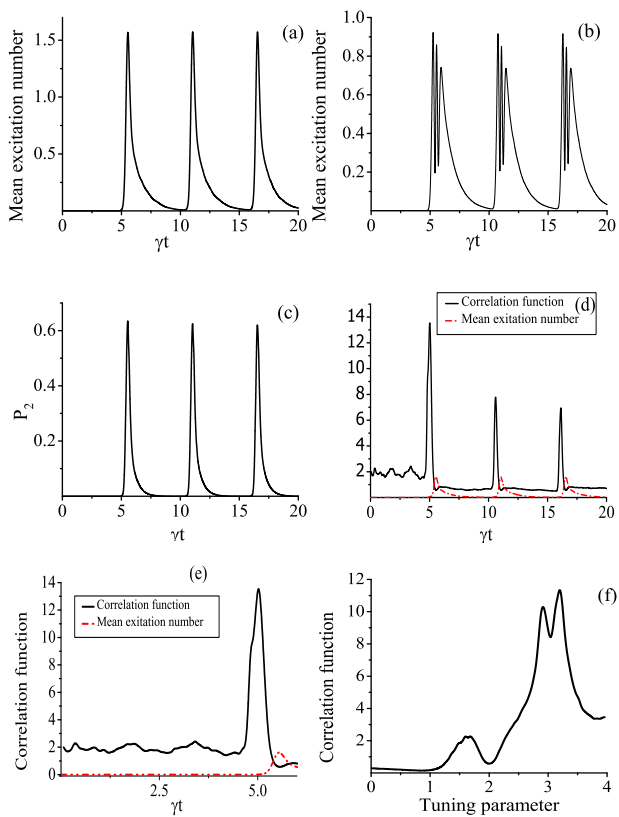}
\caption{The regimes of single-photon and two-photon blockades. The mean photon numbers versus dimensionless time-interval for $k=2$ (a) and $k=1$ (b) that repeat the Gaussian pulses and involve also decay effects; time-evolution of the population $P_2$ of $|2\rangle$ Fock state, for $k=2$, versus dimensionless time (c); the second-order photon correlation function versus the dimensionless time, for $k=2$ (d), (e); the second-order photon correlation function for time-intervals at the peak values of the mean photon numbers versus the tuning parameter (f). The parameters are:
$\chi /\gamma = 30$, the maximum amplitude of pump field $\Omega/\gamma = 12$, the mean number of reservoir photons $N=0.01$,
$\tau = 5.5{\gamma}^{-1}$, $T=0.4{\gamma}^{-1}.$}
\label{stb}
\end{figure}
\begin{figure}
\includegraphics[width=8.6cm]{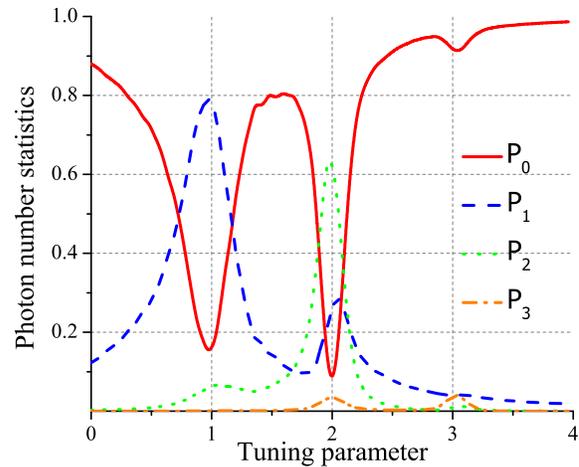}
\caption{The regimes of single-photon and two-photon blockades. The populations $P(n)$ for $n=0, 1, 2, 3$, for time-intervals at the peak values of the mean photon numbers versus the tuning parameter. The parameters are: $\chi /\gamma = 30$, the maximum amplitude of pump field $\Omega/\gamma = 12$, the mean number of reservoir photons $N=0.01$,
$\tau = 5.5{\gamma}^{-1}$, $T=0.4{\gamma}^{-1}$.}
\label{fed22}
\end{figure}

To observe photon blockade we also turn to calculation of the normalized second-order correlation function for zero delay time $ g^{(2)}$ that is defined as:
\begin{equation}
g^{(2)}(t)=\frac{\langle a^{\dagger}(t)a^{\dagger}(t)a(t)a(t)\rangle}{(\langle a^{\dagger}(t)a(t)\rangle)^2}.
\end{equation}
The results are depicted in Fig. \ref{spb}(b, c, d). The nonstationary
correlation function versus dimensionless time intervals is shown in Fig.
\ref{spb}(b) with the curve of mean photon number, while the result describing
the photon correlation for  short time intervals is presented in Fig.
\ref{spb}(c). As we see, for short time intervals $t = (0-5){\gamma}^{-1}$ only
thermal cavity photons are involved and the time dependence of $ g^{(2)}(t)$
in this area describes photon correlation of thermal bath at the averaged level $ g^{(2)}=2$.
 During the pulses, if $P_1$ reach to the  maximum, the probability of generation of
a second photon is suppressed. In this case oscillatory mode acquired
sub-poissonian statistics with the second-order correlation function $ g^{(2)}<
1$. In this way, the transition to PB is confirmed by the observation of the
second-order correlation function that shows the transition from photon
bunching to antibunching. The second-order correlation function versus the
tuning parameter is plotted in Fig. \ref{spb}(d). As calculations show,
$g^{(2)}=0.1$ for $k=1$ in accordance with the result that for $|1\rangle$ Fock
state the normalized second-order correlation function is zero. The correlation
function displays the peak at $k=1.6$. This result corresponds to the photon
bunching with the level of mean photon number $\langle n\rangle=0.38$. Then the
second-order correlation function increases with increasing of the tuning
parameter $k$, however, for negligible levels of the mean photon number. 

\subsection{Two-photon and three-photon blockades}

In this subsection we analyse the multiphoton blockade on the base of
photon-number effects. We will demonstrate how to chose the frequency of the
driving field, the parameters of Gaussian pulses and the parameter of
nonlinearity in order to realize effective populations of two-photon and
three-photon states that indicate multiphoton blockades. It is obvious that for
multiphoton blockade the regimes of photon state excitations with high levels
of photon number in compared with the case of  single-photon state excitation  should be considered.
Such regimes imply a larger peak strength of driving field or/and a comparative
small nonlinear parameter. We present below the results for two possible
operational regimes.

The typical results indicating the two-photon blockade are depicted in Fig.
\ref{stb} and Fig. \ref{fed22}. The time evolution of the mean photon number versus
dimensionless time is depicted in Fig. \ref{stb}(a) for the tuning parameter
$k=2$, (for the detuning $\Delta/\gamma=-\chi/\gamma$), corresponding to
two-photon resonant excitation; the analogous result for one-photon excitation,
$k=1$,  is shown in Fig. \ref{stb}(b).
This behavior repeats the periodicity of the driving pulses at the over
transient regime. The analogous time-evolution is observed for the population
$P_2$ at $k=2$ (see, Fig. \ref{stb}(c)).
This regime is favorable for the selective excitation of two-photon number state, if the detuning $\Delta/\gamma=-\chi/\gamma$. In this case the population $P_2=0.61$ takes place due to two-photon excitation at the frequency $\omega_2=\omega_0+ \chi $, for the tuning parameter $k=2$ (see, Fig. \ref{fed22}), while the population of one-photon state has a comparative small value $P_1=0.28$ . The peculiarity of this regime is that both one-photon blockade (for $k=1$) and two-photon blockade (for $k=2$) are effectively realized for the same parameters: $\chi /\gamma = 15$, 
 $\Omega/\gamma = 6$, $\tau = 5.5{\gamma}^{-1}$, $T = 0.4{\gamma}^{-1}$. It should be mentioned that population 
$P_2$ comes over cw limit for the population of $|2\rangle$ Fock state and (that it is remarkable) for the population of $|1\rangle$ state in cw regime.

\begin{figure}
\includegraphics[width=8.6cm]{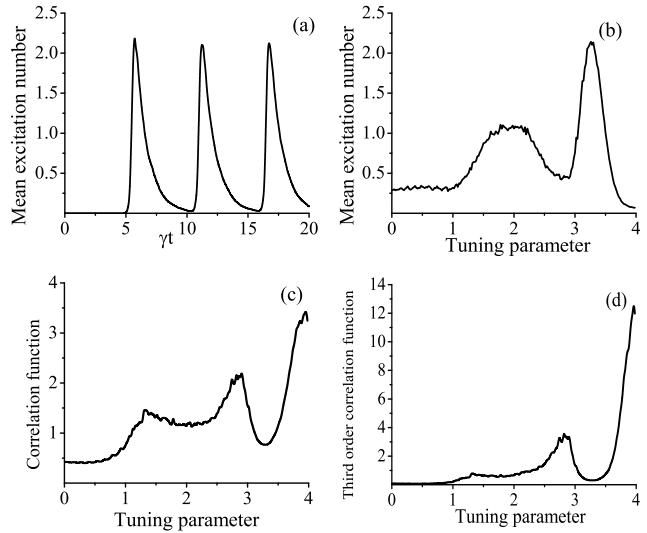}
\caption{The regime of three-photon blockades. The mean photon number versus dimensionless time, for $k=3$ (a); the maximal peak values of the mean photon numbers depending on the tuning parameter (b); the second-order correlation function at the peak values of the mean photon numbers versus the tuning parameter (c); the third-order correlation function at the peak values of the mean photon numbers versus the tuning parameter (d). The parameters are:
$\chi /\gamma = 11$, the maximum amplitude of pump field $\Omega/\gamma = 12$, the mean number of reservoir photons $N=0.01$,
$\tau = 5.5{\gamma}^{-1}$, $T=0.4{\gamma}^{-1}.$}
\label{fed31}
\end{figure}

In Fig. \ref{stb}(e) the normalized second-order correlation function versus
dimensionless time is shown. The time-evolution involves photon correlation of
thermal bath for time-interval that corresponds to resonator that involves
only photon of thermal bath. Then, the peaks of $ g^{(2)}$ take place at
time-intervals at the front of pulses. Here quantum statistics is formed, but
the mean photon number is negligible, therefore the peak in the correlation
function is observed. At definite time intervals corresponding to peaked
values of the mean photon number the correlation function equals to $
g^{(2)}=0.6$. These values of the correlation function are investigated in
depending on the detuning (the parameter of tuning) and are shown in Fig. \ref{stb}(f). As we see, for $k=1$ the effective population of single-photon number
state is observed, $P_1=0.79$ (see Fig. \ref{fed22}), and photon antibunching is realized
$ g^{(2)}=0.6$. At $k=2$, for which the maximal $P_2=0.61$ population is
realized, the correlation function equals to $ g^{(2)}=0.6$ for the mean
photon numbers $\langle n\rangle=1.52$. Note, for the comparison that for
$|2\rangle$ pure photon-number state the normalized second-order photon
correlation function equals to 0.5. In this regime the photon bunching, $
g^{(2)}=2.2$, occurs for $k=1.6$. In the vicinity of $k=3$ the large level of $
g^{(2)}$ is explained by the small level of the mean photon number (see, Fig. \ref{fed31}(a)).

\begin{figure}
\includegraphics[width=8.6cm]{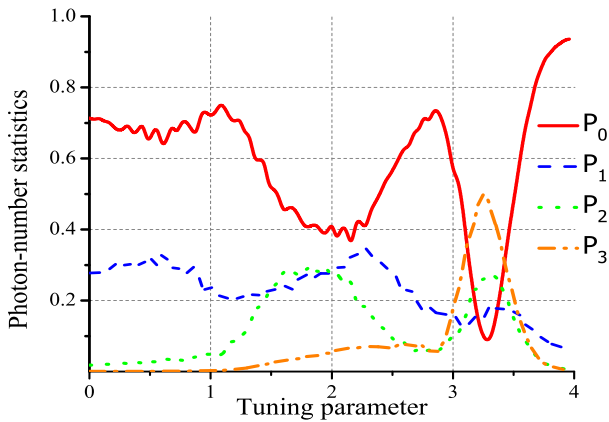}
\caption{The regime of three-photon blockades. The populations $P(n)$ for $n=0, 1, 2, 3$, for time-intervals at the peak values of the mean photon numbers versus the tuning parameter. The parameters are:
$\chi /\gamma = 11$, the maximum amplitude of pump field $\Omega/\gamma = 12$, the mean number of reservoir photons $N=0.01$,
$\tau = 5.5{\gamma}^{-1}$, $T=0.4{\gamma}^{-1}.$}
\label{fed32}
\end{figure}

In the end of this subsection we shortly discuss three-photon blockade. For
observation of the selective excitation of three-photon states we need the
operational regimes with more large levels of photon number than has been used
for the previous cases. The typical results for this regime are plotted in Fig.  \ref{fed31} and Fig. \ref{fed32}. Time-evolution of the mean photon number versus
dimensionless time is plotted in Fig. \ref{fed31}(a). This result shows the
comparative large peak value of the mean photon number for the parameters
$\chi /\gamma = 11$ and $\Omega/\gamma = 12$. In this case,  Fig. \ref{fed32}
demonstrates large population $P_3=0.48$ of $|3\rangle$ Fock state for the tuning parameter
$k=3.28$. Thus, maximal population $P_3$ takes place for near to the
resonance frequency $\omega=\omega_0+2.28\chi$. 

The maximal values of the mean photon numbers for time-interval during pulses
versus the tuning parameter are depicted in Fig. \ref{fed31}(b). This curve displays
two-peak structure of the mean photon number. The first peak occurs at $k=2$
which corresponds approximately to the mean photon number $\langle
n\rangle=P_1+2P_2+3P_3$=1.3 in accordance with Fig. \ref{fed22}. The second
characteristic double peak is at $k=3.28$ and approximately equals to $\langle
n\rangle=P_1+2P_2+3P_3$=0.14+0.56+ 1.44.

It is interesting to analyze selective excitation of three-photon state in the
framework of the second-order and third-order correlation functions. The
results of calculations are depicted in Fig. \ref{fed31}(c) and Fig. \ref{fed31}(d). As we see, for
the case of the maximal population $P_3=0.48$ of $|3\rangle$ Fock state, if
$k=3.28$ and the mean photon number is $\langle n\rangle=2$ , 
the second-order correlation function is $g^{(2)}=0.75$, (note, that for the pure $|3\rangle$ state $g^{(2)}=2/3$ in accordance with the result obtained), while the
third-order correlation function reaches to minimal value $g^{(3)}=0.32$. It should be mentioned, that the last result is in accordance with the result for pure 
$|3\rangle$ Fock state, $g^{(3)}=2/9$.

We emphasize that results of this subsection indicate the increasing of the
populations of two-photon and three-photon states of Kerr nonlinear resonator
in the pulsed regime in comparison with the analogous results observed in the
case of cw excitation (see, \cite{Mir} and the results depicted in Fig. \ref{monodelta} and
Fig. \ref{monochi}). This observation indicates on the efficiency of multiphoton blockades
for the pulsed regimes.

\section{Production of multi-photon states in the pulsed regimes}

Recently, it has been demonstrated that formation of single-photon states can
be realized in a cavity strongly coupled with emitter operated in the pulsed
regime. In this way the photon-number effects and phase-space characteristics
of photonic states can be controlled by choosing the parameters of driving
pulses \cite{A}, \cite{gev2}. In this section, we investigate production of
two-photon and three-photon light states considering mainly the selectivity of the excitation of
these states by driving pulses  and their phase-space properties.

\begin{figure}
\includegraphics[width=8.6cm]{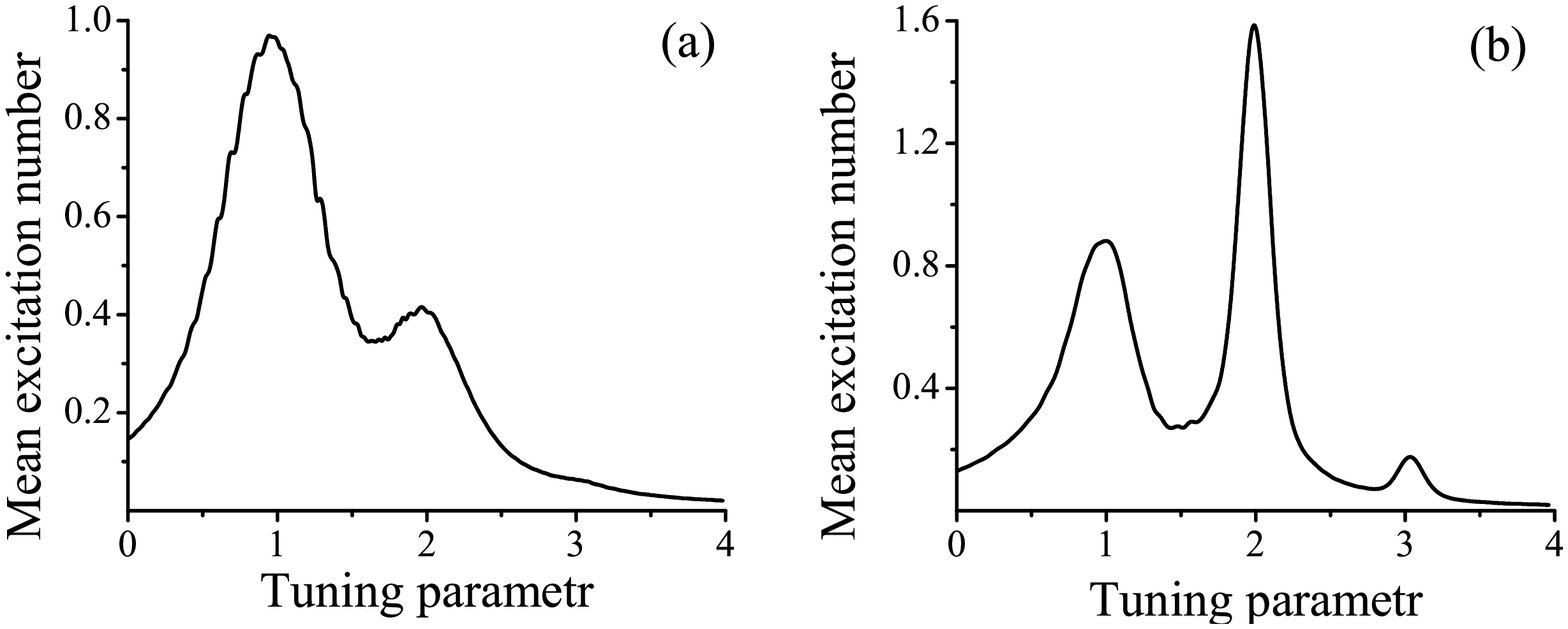}
\caption{The maximum values of the mean photon numbers  depending on the tuning  parameter. The parameters are:
$\chi /\gamma = 15$, $\Omega/\gamma = 6$, $\tau = 5.5{\gamma}^{-1}$, $T=0.4{\gamma}^{-1}$ (a); $\chi /\gamma = 30$, $\Omega/\gamma = 12$,
$\tau = 5.5{\gamma}^{-1}$, $T=0.4{\gamma}^{-1}$ (b). The mean number of reservoir photons $N=0.01$.}
\label{fed1}
\end{figure}

At first, we turn to the regime of single-photon blockade shown in Fig. \ref{spb} and
Fig. \ref{pns} considering maximal values of the mean photon number for time-interval
during pulses depending on the tuning parameter. The result of
calculations are depicted in Fig. \ref{fed1}(a) where we plot the dependence of the
mean photon number for different detunings. This curve shows a large spectral
width of generation with the maximum at $\Delta=0$. The peaked intensity is
approximately $\langle n\rangle=P_1+2P_2$=0.82+ 0.18 with the populations have
been obtained in Fig. \ref{pns}. 

Now, we consider production of two-photon states by selective two-photon
excitations in Kerr nonlinear resonator. In order to demonstrate the process we
turn to the regime of two-photon blockade (see, the results depicted in Fig. \ref{stb}
and Fig. \ref{fed22}). For these parameters the population of two-photon number state
reaches $P_2=0.61$ for the detuning $\Delta/\gamma=-\chi/\gamma$.
The curve of the maximal values of the mean photon numbers for these
parameters are plotted in Fig. \ref{fed1}(b) in dependence from the  detuning values. As we see,
the mean photon number displays two-peak structure. The first peak occurs at
$k=1$ which corresponds approximately to the mean photon number $\langle
n\rangle=P_1+2P_2$=0.82+0.18. The second characteristic double peak is at $k=2$
and approximately equals to $\langle n\rangle=P_1+2P_2$=0.3+ 1.22.

To obtain a complete description of the system we turn to calculation of the Wigner function in phase-space. The numerical results are given for the regime used above that relies to effective selective excitation of two-photon states, for $k=2$.

\begin{figure}
\includegraphics[width=8.6cm]{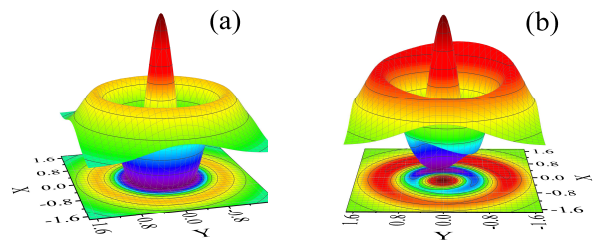}
\caption{The regime of selective two-photon excitation. The Wigner function for
the pure $|2\rangle$ state (a) ; the Wigner function of cavity mode for time-interval corresponding to the maximal
 population of Fock state $|2\rangle$  (b). The negative regions
of the Wigner functions are indicated in black. The parameters are: $\chi
/\gamma = 15$, the maximum amplitude of pump field $\Omega/\gamma = 12$, the
mean number of reservoir photons $N=0.01$, $\tau = 5.5{\gamma}^{-1}$,
$T=0.4{\gamma}^{-1}$.}
\label{stp}
\end{figure}

The Wigner function of intracavity mode for time-intervals corresponding to the
maximal $P_2=0.61$ population is shown in Fig. \ref{stp}(b) with the Wigner function
of pure $|2\rangle$ state, plotted in Fig. \ref{stp}(a)) for comparison.  It is easy to realize
that this Wigner function displays a ring signature with the center at $x = y =
0$ in the phase space and the good agreement in shape takes place with the Wigner function of  pure $|2\rangle$ state.

\begin{figure}
\includegraphics[width=8.6cm]{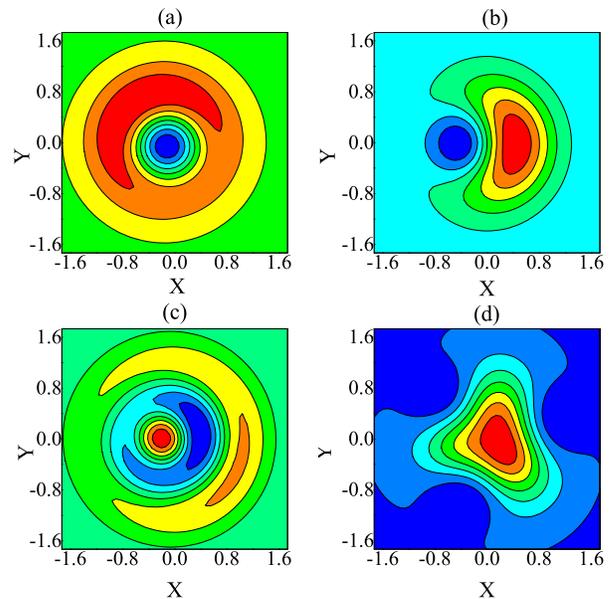}
\caption{ The contour plots of the Wigner functions for the  various selective excitations: $k=1$ (a); $k=1.24$  (b), where the  populations of $|0\rangle$ and $|1\rangle$ are crossing each other (see, Fig. 8); $k=2$ (c); $k=3$ (d). The negative regions of the Wigner functions are indicated
in black. The parameters are:
$\chi /\gamma = 15$, the maximum amplitude of pump field $\Omega/\gamma = 12$, the mean number of reservoir photons $N=0.01$,
$\tau = 5.5{\gamma}^{-1}$, $T=0.4{\gamma}^{-1}$.}
\label{fed2}
\end{figure}

Fig. \ref{fed2} shows comparative analysis of the contour plots of the various Wigner
functions corresponding to the definite detuning values that determine various
resonant one-photon and multi-photon transitions.
As we see, for the regime of one-photon excitation, $\Delta=0$, on the
frequency $\omega_1=\omega_0$ the Wigner function (Fig. \ref{fed2}(a)) displays good
agreement with the Wigner function of pure  $|1\rangle$ single-photon state. In
Fig. \ref{fed2}(b) the result for $k=1.24$, i.e. for $\Delta/\gamma=-0.24\chi/\gamma$ is
presented. In this case, the populations of vacuum state and single-photon
state are crossing and are approximately equal (see, Fig. \ref{fed22}) thus, the Wigner
function seems to be closest to the Wigner function of pure superposition
state $\Psi\rangle=\frac{1}{\sqrt{2}}(|0\rangle-|1\rangle)$. For $k=2$, the contour plot depicted in Fig. \ref{fed2}(c) corresponds to the
Wigner function of Fig. \ref{stp}(b). In Fig. \ref{fed2}(d) the contour plot for the case of
near the resonant three-photon excitation at $k=3.28$ is depicted. This Wigner function shows three-phase symmetry and also indicates
the regions of quantum interference in the contour plot as the negative regions
in black. Note that three-fold symmetry of the Wigner function
and interference pattern has been demonstrated for the direct
three-photon down-conversion in $\chi^{(3)}$ media \cite{knight} and also in details for production of three-photon states in \cite{three1}, \cite{three2}.

\section{Conclusion}

We have investigated photon-number effects in multiphoton blockade for Kerr nonlinear dissipative resonator driven by sequence of Gaussian pulses. This model can be realized at least with two physical systems: a qubit off-resonantly coupled to a driven cavity or superconducting devices based on Josephson junction. We consider the cases of strong nonlinearity with respect to the rate of damping of
the oscillatory mode. In these regimes the oscillatory energy levels are well resolved, and spectroscopic selective excitation of transitions between Fock states is possible by tuning the frequency of driving field.

 Comparing the results for continuous wave- and pulsed- operational regimes of Kerr nonlinear resonator we demonstrate that the larger photon-number populations of the resonator can be reached if shaped pulses are implemented. Thus, we have shown that optimized one-photon and multi-photon blockades can be obtained by adequately choosing the duration of pulses and the time intervals between them. 

Consideration of the Kerr nonlinear dissipative resonator under cw driving has been done on the framework of the potential solution of the Fokker-Planck equation for the quasiprobability distribution function in complex
P-representation. In this way, we strongly demonstrated that for the case of cw
pumping of the Kerr-nonlinear resonator there are fundamental limits on the
populations of lower photonic-number-states (n=0, 1, 2, 3). Particularly, the
maximal value of $|1\rangle$ state population is limited by $P_1=0.5$ for all
values of the system parameters. 

More importantly, considering Kerr-nonlinear resonator driven by a sequence of
Gaussian pulses, we have demonstrated the regimes in which the larger
photon-number populations of the resonator beyond cw limits can be obtained.
This improvement is due to a quantum control of decoherence in Kerr-nonlinear
resonator by applying of suitable tailored e.g. pulses. Particularly, the
typical results for $P_1$ population overcome 0.8 value. 

We analyse photon-number effects and investigate in details the photon-number correlation effects on base of the second-order and third-order correlation functions for the case of selective excitations of intracavity mode. Specifically we have calculated the Wigner function in phase-space for the regimes of multi-photon excitations of mode.

\begin{acknowledgments}
We acknowledge support from the Armenian State Committee of Science, the
Project No.13-1C031. G. Yu. K. acknowledges discussions with I. A. Shelykh.
\end{acknowledgments}

\end{document}